\documentclass[aps,floatfix,
	prl,
	showpacs,twocolumn,groupedaddress,superscriptaddress]{revtex4}
\usepackage{graphicx}%
\usepackage{dcolumn}
\usepackage{amssymb,amsmath}
\usepackage{bm}
\usepackage[usenames]{color}

\begin{document}

\title{Turbulence in a Bose-Einstein Condensate of Dipolar Excitons\\ in Coupled Quantum Wells}

\author{O.~L.~Berman}
\affiliation{Physics Department, New York City College of Technology, CUNY, Brooklyn, NY 11201, USA}

\author{R.~Ya.~Kezerashvili}
\affiliation{Physics Department, New York City College of Technology, CUNY, Brooklyn, NY 11201, USA}

\author{G.~V.~Kolmakov}
\affiliation{Physics Department, New York City College of Technology, CUNY, Brooklyn, NY 11201, USA}

\author{Yu. E. Lozovik}
\affiliation{Institute of Spectroscopy RAS, Troitsk, Moscow region, 142190, Russia}


\begin{abstract}
The nonlinear dynamics   of a Bose-Einstein
condensate (BEC)  of dipolar excitons trapped in an external confining
potential in coupled quantum wells is analyzed. It is demonstrated that under typical experimental conditions
the dipolar excitons BEC can be described by a generalized
Gross-Pitaevskii equation with the local interaction between the
excitons, which depends on the exciton distribution function. 
It is shown that, if the system is pumped at sufficiently high frequencies, a steady turbulent state 
can be formed.
\end{abstract}


\maketitle

In the last decade, the nonlinear dynamics of the excitations in 
semiconductor heterostructures coupled with laser radiation attract attention because of promising potential 
applications in electronics and photonics, including, e.g., design of thresholdless lasers, 
optical computing and quantum computing \cite{Liew:08,Snoke:09,Amo:10}. 
One of the examples of the systems where the excitations demonstrate essentially quantum behavior is a
Bose-Einstein condensate (BEC) of excitons in semiconductors 
\cite{Lozovik:75,  Butov:94,  Zhu:95, Snoke:02, Butov:12}. 
The dynamics of dipolar excitons formed by spatially separated charges in coupled quantum wells (CQWs)  
in semiconductor heterostructures at helium temperatures  attract attention due to 
the relatively long exciton lifetime compared to excitons in a single quantum well \cite{Snoke:02,Butov:94}. 
Recently, it was predicted that dipolar excitons can form a superfluid  in a two-layer 
graphene \cite{Berman:11}.

In this paper we focus on non-stationary dynamics in a dipolar exciton BEC in QCWs in an in-plane 
trapping potential. The  trapping potential is essential for the condensate formation 
at finite temperatures and it can be formed in GaAs structures by applying inhomogeneous 
stress 
\cite{Snoke:02}, static electric as well as magnetic field or laser radiation 
(see \cite{Butov:12} and references therein).  
The dynamics of dipolar excitons is complicated
due to long-range, $\propto 1/r^3$,  interaction in the system. 
In the present work, we demonstrate that under certain conditions the interactions in a  dipolar 
exciton BEC can be considered as {\it local interactions}. 
As a result, the dynamics of the BEC can be effectively described by the 
generalized Gross-Pitaevskii equation (GPE).

To characterize the non-stationary behavior of the dipolar exciton BEC we perform numerical 
simulations of the system where the resonant driving in the low- or high-frequency spectral domains is 
present. It was found that the dynamics of 
the system in these two cases are essentially different: for the low-frequency driving, 
the BEC spatial distribution tends to a stationary bell-like shape described well by the parabolic 
Thomas-Fermi profile \cite{Dalfovo:99} whereas for the high-frequency driving, the BEC demonstrates 
long-lasting, non-stationary oscillations. 
This oscillatory state is somewhat similar to the wave turbulence state observed 
earlier in spatially restricted nonlinear superfluid systems \cite{Ganshin:08,Abdurakhimov:10}.

{\it Local approximation for a dipolar exciton BEC.}  At temperatures 
much lower than the BEC transition temperature,  the dynamics of a dilute dipolar exciton condensate is 
described by the generalized two-dimensional Gross-Pitaevskii equation 
\vspace{-0.1cm}
\begin{eqnarray}
     & & i \hbar \frac{\partial \Psi(\bm{r},t) }{\partial t} =   - \frac{\hbar^{2}}{2 m_{\rm ex}}  
        \Delta \Psi(\bm{r},t)  +  V(\bm{r})\Psi(\bm{r},t) + \Psi(\bm{r},t)  \nonumber \\ 
    & & \times \int  d^2 \bm{r}^{\prime} |\Psi (\bm{r}^\prime,t)|^{2}U(\bm{r} - \bm{r}^\prime) + 
        i\hbar \left( \hat{R} - \gamma \right) \Psi (\bm{r},t) . \label{gpnl}
\end{eqnarray}
In Eq.\ (\ref{gpnl}) $\Psi(\bm{r},t)$ is the exciton condensate wave function, $m_{\rm ex}$ is the exciton mass,
$V(\bm{r}) = \alpha r^2/2$ is an external parabolic trapping potential,  
$\alpha$ is the trapping potential strength,  
$U(\bm{\rho}) = {e^2 D^2 / \varepsilon \rho^3}$ is the pairwise exciton interaction potential,  
$e$ is the electron charge, $D$ is the inter-well distance, $\varepsilon$ is the dielectric 
permittivity of the semiconductor,  $\rho = |\bm{r} - \bm{r}^{\prime}|$ is the distance between the excitons,
and $\gamma=1/2 \tau_{\rm ex}$, where $\tau_{\rm ex}$ is the exciton lifetime.
Below an isotropic trap is considered. However, this is not a restriction of the model and we can also 
consider anisotropic traps with different trapping potential strengths in $x$ and $y$ directions.
The last term in Eq.\ (\ref{gpnl}) describes  creation of the excitons due to the interaction 
with the pumping laser radiation and the exciton decay similarly to that used in earlier works \cite{Keeling:08,Amo:09}. 
However, to capture the resonant pumping  in a given spectral domain, we introduce 
a linear operator $\hat{R}$ instead of the direct driving \cite{Amo:09} or 
the homogenous, frequency-independent growth  increment \cite{Keeling:08}. 
The $\hat{R}$ operator is defined via its matrix elements in a functional 
basis that enter into the system of equations (\ref{eqa}) given below. 

\definecolor{darkgreen}{rgb}{.0,0.5,.0} 
\begin{figure}[t] 
\includegraphics[width=58.mm]{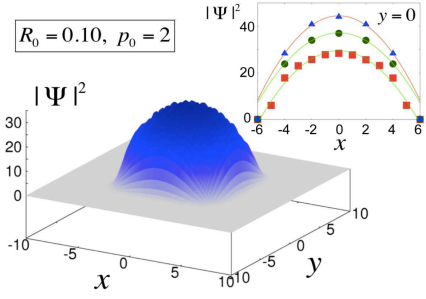}
\vspace{-3.mm}
\caption{\label{fig1} Steady-state spatial distribution for the dipolar exciton BEC density in a trap for the 
low-frequency driving at $R_0=0.1$. 
Inset shows the evolution of the exciton density profile with rising the pumping rate $R_0$:  
 0.1 ({\color{red} $\blacksquare$}), 0.3 ({\color{darkgreen} \Large $\bullet$}), 
and 0.5 ({\color{blue} $\blacktriangle$}).
The curves represent the Thomas-Fermi distribution function.} 
\end{figure}

The integral term in Eq. (\ref{gpnl}) diverges at small
distances $\rho$, that corresponds to known divergence of the scattering 
amplitude of the dipolar excitons \cite{Lozovik:75}. 
To regularize the integral we introduce a cutoff  distance 
$r_0 = (\varepsilon \mu / e^2 D^2)^{1/3}$ defined by the equation 
$U(r_0) = \mu$, where $\mu$ is the chemical potential in the exciton system. 
This is equivalent to the cutoff of the exciton energies 
at the chemical potential $\mu$. In this case,  the function 
$|\Psi (\bm{r}^\prime,t)|^{2}$ can be expanded in a  Taylor series over  $\rho$ and 
the integral term reads  $\int_{\rho \geq r_0}  d^2 \bm{r}^{\prime} |\Psi (\bm{r}^\prime,t)|^{2} U(\rho) = 
g |\Psi (\bm{r},t)|^{2} (1 + O(r_0/a))$, where $g = 2 \pi e^2 D^2/\varepsilon r_0$, 
$O(r_0/a)$ denotes the terms of $(r_0/a)$ order, 
and $a \sim \Psi (\bm{r},t) / |\nabla  \Psi (\bm{r},t)|$ is the characteristic length,
at which the condensate wave function changes significantly.
For the low-frequency resonant driving $a$ coincides by the 
order of magnitude with the size of the exciton cloud  $d \sim 10$ $\mu$m \cite{Snoke:02,Butov:12}. Hence,
the used approximation for the integral term is valid if $r_0 \ll 10$ $\mu$m. For the high-frequency driving, 
$a$ can be smaller than $d$ and is defined by the driving frequency. The estimates  for the experimental conditions
\cite{Snoke:02,Butov:12} show that $r_0/d \leq 0.04$ and therefore, the corrections $\sim (r_0/a)$ are negligible.

Thus, under the experimental conditions \cite{Snoke:02,Butov:12} the dynamics of the dipolar exciton BEC is described by the equation
\vspace{-0.2cm}
\begin{eqnarray}
      & & i \hbar \frac{\partial \Psi(\bm{r},t) }{\partial t} =    - \frac{\hbar^{2}}{2 m_{\rm ex}}  
        \Delta \Psi(\bm{r},t)  +  V(\bm{r})\Psi(\bm{r},t)  \nonumber \\ 
    & & + {g} \Psi (\bm{r},t) |\Psi (\bm{r},t)|^{2} + 
        i \hbar \left( \hat{R} - \gamma \right) \Psi (\bm{r},t).  \label{gpp}
\end{eqnarray}
Eq. (\ref{gpp}) coincides with the ``traditional'' GPE, in which creation and decay of the
particles are taken into account. However,
in Eq. (\ref{gpp}) the exciton interaction strength $g$ depends on the chemical potential of 
the system and thus, on the density of the exciton BEC. In what follows we incorporate the equation for
the chemical potential in the BEC determined for a dilute gas in a trap \cite{Dalfovo:99},
\vspace{-0.2cm}
\begin{equation} 
  \mu = (H^{(2)} + 2 H^{(4)}) / N, \quad 
  H^{(4)} = {g \over 2}  \int d\bm{r} |\Psi (\bm{r},t)|^{4}, \label {eqmu}
\end{equation}
\vspace{-0.7cm}
\[
   H^{(2)} =  \int d\bm{r} \left( {\hbar^2 \over 2 m_{\rm ex}} | \nabla \Psi (\bm{r},t)|^{2} +  
   |\Psi (\bm{r},t)|^{2} V(\bm{r})\right) ,
\] 
where $N = (1/2) \int d\bm{r} |\Psi (\bm{r},t)|^{2}$ is the total number of the 
excitons in the BEC, and $H^{(2)}$ and $H^{(4)}$ are the quadratic and fourth-order
terms in the Hamiltonian of the system, respectively. 

\begin{figure*}[t] 
\includegraphics[width=65.mm]{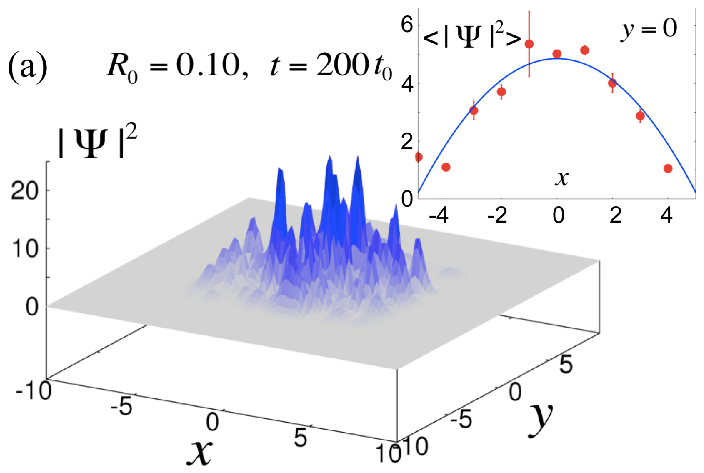}\hspace{1.0cm}
\includegraphics[width=55.mm]{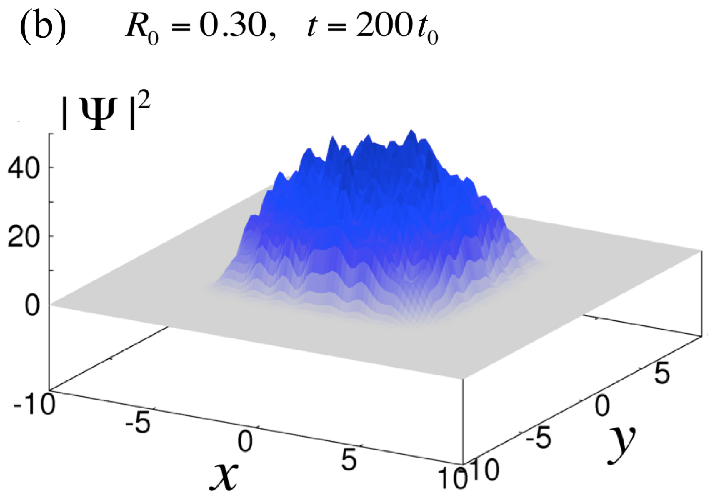}
\vspace{-3.mm}
\caption{\label{fig2}  The exciton density profiles at  $t=200 t_0$ for (a) $R_0 = 0.1$ 
and (b) $R_0 = 0.3$ for the  high-frequency driving at $p_1 = 4$, $p_2 = 6$. Inset on frame (a) shows the exciton 
density plotted at $y=0$ and averaged over the time period $50 t_0 < t < 200 t_0$ and 
three independent runs (points). Curve shows the fitting  by the Thomas-Fermi distribution.}
\end{figure*}

{\it Exciton BEC formation.}  To investigate the nonlinear dynamics of a dipolar exciton BEC,  we solve the 
system (\ref{gpp}), (\ref{eqmu}) with the help of the spectral representation that results in the following
 system of equations 
 \vspace{-0.2cm}
\begin{equation}
 {\partial A_{\bm{n}} \over \partial t} = -i g \sum_{\bm{m},\bm{k},\bm{l}}  W_{\bm{n},\bm{m},\bm{k},\bm{l}}
 A_{\bm{m}}^*A_{\bm{k}}  A_{\bm{l}} +  (R_{\bm{n}} - \gamma ) A_{\bm{n}}, \label{eqa}
 \end{equation}
where  $A_{\bm{n}}(t)$ are  time-dependent spectral amplitudes,
$W_{\bm{n},\bm{m},\bm{k},\bm{l}} $ is the matrix element of the interaction term $H^{(4)}$, 
and $R_{\bm{n}}$ is the matrix element of the $\hat{R}$ operator. 
$A_{\bm{n}}(t)$  are the coefficients of the expansion of the condensate wave function,
$ \Psi (\bm{r},t) = \sum_{\bm{n}} A_{\bm{n}}(t) \varphi_{\bm{n}}(\bm{r}), $
where the  basis functions $\varphi_{\bm{n}}(\bm{r})$ are the eigenfunctions 
of the Hamiltonian  for a single particle in a parabolic potential \cite{Landau:77}, and 
$\bm{n} = (n_x,n_y)$ is the two dimensional index.
The Schr\"{o}dinger equation for the basis functions coincides with the linear, 
Hermitian part (i.e. the first two terms in the r.h.s.) of Eq. (\ref{gpp}).
This choice allows us to capture the  behavior of the condensate 
wave function by taking into account a relatively small  number of  terms in the expansion for $\Psi (\bm{r},t)$ and hence, 
obtain a good numerical  convergence. 
In the simulations, the length and time 
are expressed in the oscillatory units $\ell_0 = (\hbar^2 / \alpha m_{\rm ex})^{1/4} = 0.9$ $\mu$m, 
$t_0 = (m_{\rm ex}/\alpha)^{1/2} = 1.6$ ns calculated for the external trapping potential 
at $\alpha = 50$ eV/cm$^2$ and the exciton mass $m_{\rm ex} = 0.22m_0$, where $m_0$ is the free electron mass. 
The initial conditions were chosen in the form of quasi-equilibrium distribution 
$A _{\bm{n}} (0) = [T / (\mu_0 + n_x + n_y + 1)] ^{1/2} \exp(i\phi_{\bm{n}})$
with the dimensionless  temperature $T = 0.1$,  the dimensionless chemical potential 
$\mu_0 = 1$, and random phases $\phi_{\bm{n}}$. We found that the results only weakly depend on the choice 
of the $T$ and $\mu_0$ constants.  We numerically integrated the system of equations (\ref{eqa}) 
with the 4$^{\rm th}$ order Runge-Kutta scheme  on a graphical processing unit NVIDIA Tesla S2050.

We consider two cases: (a) the exciton system is pumped at low spectral modes, 
 {\it i.e.}, at frequencies comparable with the fundamental frequency in the parabolic trap, 
 $\omega_0 = t_0^{-1}$; (b) the system is pumped
in the high frequency spectral domain  $\omega > \omega_0$. 
It is worth noting that, in experiments the ratio of the driving frequency
and the fundamental frequency of the trap may be tuned by changing both the laser pumping frequency and 
the trapping potential strength $\alpha$ \cite{Snoke:02,Butov:12}.

For the low-frequency driving, we set $R_{\bm{n}} = R_0$ at $n \equiv (n_x^2 + n_y^2)^{1/2} \leq p_0$ 
where $p_0$ is a given cut-off 
and $R_{\bm{n}} = 0$ otherwise. Below, we show the results obtained for $p_0=2$. The driving was applied at $t=0$. Initially,  
the system exhibits transient oscillations.  At a later time, the oscillations were damped and the system relaxed to
a stationary state, in which the spectral amplitudes $A_{\bm{n}}$ tend to constant values. 
The time required to approach the stationary state was equal to $t \approx 50 t_0 \approx 80$ ns 
for a low pumping rate $R_0 = 0.1$ and it decreases to $t \approx 20 t_0 \approx 32$ ns for a high pumping rate $R_0 = 0.6$. 
We also observed that the initial exciton distribution relaxed to zero if the pumping rate was less that the threshold 
value $R_0 \approx 0.03$. The presence of the finite excitation threshold is in agreement with the results of previous 
simulations of the polaritons  dynamics \cite{Keeling:08}.
Figure \ref{fig1} shows the stationary non-uniform spatial distribution of the exciton BEC density  at 
$t = 200t_0$ for $R_0 = 0.1$. It is seen in Fig. \ref{fig1} that an exciton ``cloud'' of size $\sim 6-8$ $\ell_0$ 
that is, $\sim 5 - 7$ $\mu$m is formed at the center of the trap. Formation  of the cloud at the center of the trap 
is in agreement with the observations with indirect excitons in a trap \cite{Snoke:02} and in the BECs of trapped atomic gases 
\cite{Dalfovo:99}. As is seen in the inset in Fig.\ref{fig1}, the spatial distribution of the 
exciton in the BEC can by described by the Thomas-Fermi distribution 
$|\Psi(\bm{r})|^2 = n_0 [1 - (r/d)^2]$ ($n_0$ is the BEC density at the center of the 
trap and $d$ is the effective radius of the cloud), which is used for characterization of 
inhomogeneous atomic BECs \cite{Dalfovo:99}.   It is seen in Fig.\ \ref{fig1} that the exciton density at the 
center of the trap gradually grows with the increase of the pumping rate.

\begin{figure}[h] 
\vspace{-2.mm}
\centerline{\includegraphics[width=55.mm]{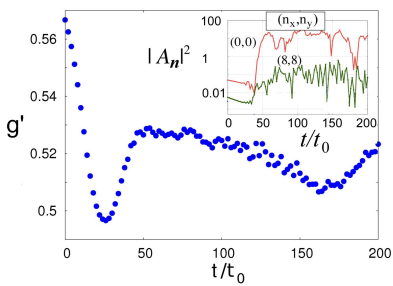}}
\vspace{-3.mm}
\caption{\label{fig7}  Dependence  of the dipolar exciton interaction 
strength on time for the high-frequency driving at $p_1 = 4$, $p_2 = 6$. 
Inset: Time dependence of the spectral amplitudes
$|A_{\bm{n}}|$ at $\bm{n} = (0,0)$ (the fundamental mode) and $(8,8)$.}
\end{figure}

When the system was driven in the high frequency spectral range ($R = R_0$ at $p_1 < n \leq p_2$,  and $p_2 > p_1 \geq 4$) 
the BEC exhibited {\it persistent, non-damped oscillations}.  
At relatively small pumping rates, $R_0 \sim 0.1$,  the spatial BEC profile at any given moment of time 
was far from the Thomas-Fermi distribution and it was represented as a series of irregular ``spikes'', as seen in Fig. \ref{fig2}a. 
Nevertheless, the  mean exciton BEC density distribution averaged over a long enough  period of time, $\langle |\Psi(\bm{r},t)|^2 \rangle$, 
was close to the Thomas-Fermi distribution, as shown in the inset in Fig. \ref{fig2}a. At larger pumping rates, 
$R_0 \geq 0.3$, the density oscillated around a mean parabolic profile, which was close to the Thomas-Fermi distribution 
(Fig. \ref{fig2}b). We emphasize that, unlike the case of the low frequency driving, the BEC oscillations were observed 
during the whole time period of the simulations $200 t_0$, 
which is much longer than the time duration of the initial transient processes. Figure \ref{fig7} shows the respective 
oscillations of the spectral harmonics, $A_{\bm{n}}$, at $t > 50 t_0$ and the variations of the 
dimensionless interaction strength, $g^{\prime} = g m_{\rm ex} / \hbar ^2$, with time. 
A decrease of both $A_{\bm{n}}$ and $g^{\prime}$ at $t<50t_0$ corresponds to 
initial relaxation after the driving was applied. 

\begin{figure}[t] 
\includegraphics[width=55.mm]{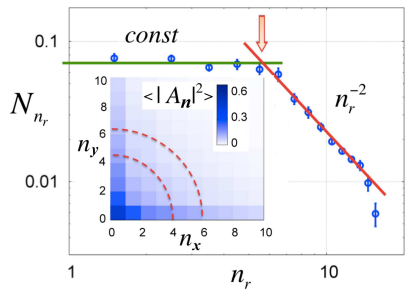}
\vspace{-0.4cm}
\caption{\label{fig6} Angle-averaged occupation number of the excitonic modes in the BEC, $N_{n_r}$ as a function of the radial 
number $n_r$, plotted in log-log scale. The data are averaged over three independent runs. 
The center of the pumping region is labeled by a vertical arrow, $\Delta n = 3$. 
The lines show a power-like distribution for $N_{n_r} = const \times n_r^{m}$ at $m = 0$ and $m = -2$.  
Inset shows the averaged spectral amplitude distribution $|A_{\bm{n}}|^2$, the dashed curves show the boundaries of 
the pumping domain.}
\end{figure}

To characterize the oscillatory regime, we plot  a two-dimensional amplitude distribution 
$\langle |A_{\bm{n}}|^2 \rangle$ averaged over time and three independent runs, 
as a function of $(n_x, n_y)$, see inset in Fig. \ref{fig6}. It is clearly seen that, 
despite the system is driven at the high-frequency range (marked by dashed curves), 
the maximum of the spectral amplitudes (deeper color) is positioned at low spectral numbers
$n \leq 2$. Therefore,  mutual interaction between different 
spectral scales is present in the system. In other words, a flux of the excitations from the region, 
at which they are generated by an external pumping, toward the low-frequency region is formed. 
The resultant state is quite similar to a wave-turbulent state known for weakly nonlinear systems 
\cite{Proment:09}. In those systems, nonlinear interaction of running waves 
resulted in formation of turbulence, which was characterized by establishing of
power-like, Kolmogorov spectra of energy and particles distributions. 
In our case, the system is not spatially homogenous and the interacting, normal variables 
are present by the oscillatory modes $A_{\bm{n}}$.
 
To further characterize the turbulent regime, we calculate the angle-averaged distribution for the spectral amplitudes,
$N_{n_r} = \sum_{n = n_r} ^  {n_r + \Delta n} \langle |A_{\bm{n}}|^2 \rangle$,
see Fig. \ref{fig6}. Formation of power-like tails of 
the distribution $N_{n_r} \propto n_r ^ {m}$  at frequencies below and above the characteristic driving frequency 
is clearly seen on the main part of  Fig.\ \ref{fig6}, that supports our conclusion on 
the establishing of the turbulent regime in the dipolar exciton BEC in CQWs. 
The presence of two different exponents $m = 0$ and $m=-2$ in the low- and high-frequency domains, respectively,
indicate that the fluxes of the energy and of the number of particles through the scales 
are simultaneously formed in the system 
(cf. theory for traditional GPE \cite{Proment:09}). An essential difference between the previous 
considerations  \cite{Proment:09} and our results is that the effective damping in the system 
produced by the exciton relaxation is finite  and cannot be disregarded for all modes. 
In effect, the fluxes of the energy and the number of particles are not completely conserved in the spectral space.  

In conclusion, we demonstrate that the dynamics of the Bose-Einstein condensate of 
the excitons in coupled quantum wells can be described 
by the generalized Gross-Pitaevskii equation with the local (contact) 
interaction, despite a long-range dipolar exciton interaction  
is present. The effective interaction strength $g$ is a function of the chemical potential of the system 
and therefore, should be determined self-consistently from the exciton distribution in the BEC. We show that,
if the system is driven by an external pumping at low frequency modes, the spatial distribution of the excitons  
in a trap is described well by a parabolic density profile formerly known for dilute atomic BECs. 
However, if the system is driven at high frequency modes, strong time-dependent density fluctuations are 
excited in the BEC, and the condensate density at any given moment of time can be far from the equilibrium 
parabolic profile.
We infer that a  turbulent state is formed in the exciton BEC in this case, which is characterized by a nonlocal 
particle balance in the system. The latter results in formation of the power-like spectra for the exciton distribution function. 
The turbulent state is somewhat similar to that recently observed in superfluid $^4$He 
\cite{Ganshin:08,Abdurakhimov:10} and proposed in 
\cite{Lvov:03} for the atomic BEC formation. The crucial difference between all 
mentioned examples and the system under consideration is that in our case the interaction strength 
between the normal modes is a function of the occupation number of the modes. 
Therefore, turbulence can not be considered as ``weak" and
it involves more complex interactions between the modes.
It is worth noting that formation of long-living nonequilibrium states was recently described for a classical system 
with long-range interaction \cite{Benetti:12}.
The results of our consideration are useful for understanding of nonlinear phenomena in low-dimensional quantum systems  including indirect excitons in CQWs \cite{Snoke:02,Butov:12}, exciton polaritons propagation  
\cite{Amo:10}, and a photon BEC in dye microcavities \cite{Klaers:12}.  


\end{document}